\documentclass{emulateapj}

\usepackage{amsmath}
\usepackage{xspace} 
\usepackage{graphicx}
\usepackage{natbib}

\shorttitle{The Stress Edge in Sgr A*}

\begin{document}

\title{Modulated X-ray Emissivity near the Stress Edge in Sgr A*}

\author{Maurizio Falanga\altaffilmark{1,2}, Fulvio
  Melia\altaffilmark{3,4}, 
Martin Prescher\altaffilmark{3}, Guillaume
  B\'elanger\altaffilmark{5}, 
Andrea Goldwurm\altaffilmark{1,6}} 

\altaffiltext{1}{CEA Saclay, DSM/IRFU/Service d'Astrophysique, 
91191 Gif-sur-Yvette, France; mfalanga@cea.fr}
\altaffiltext{2}{AIM - Unit\'e Mixte de Recherche CEA - CNRS -
  Universit\'e Paris Diderot}
\altaffiltext{3}{Physics Department and Steward Observatory, The
  University of Arizona, Tucson, AZ 85721} 
\altaffiltext{4}{Sir Thomas Lyle Fellow and Miegunyah Fellow}
\altaffiltext{5}{ESA/ESAC, Apartado 50727, 28080 Madrid, Spain}
\altaffiltext{6}{UMR Astroparticule et Cosmologie, 
10, rue Alice Domont et L\'eonie Duquet, 75005 Paris Cedex 13, France} 
\begin{abstract}

Sgr A* is thought to be the radiative manifestation of a $\sim 3.6\times 
10^6$ $M_\odot$ supermassive black hole at the Galactic center. Its 
mm/sub-mm spectrum and its flare emission at IR and X-ray wavelengths may 
be produced within the inner ten Schwarzschild radii of a hot, magnetized 
Keplerian flow. The lightcurve produced in this region may exhibit 
quasi-periodic variability. We present ray-tracing simulations to 
determine the general-relativistically modulated X-ray luminosity 
expected from plasma coupled magnetically to the rest of the disk 
as it spirals inwards below the innermost stable circular orbit 
towards the ``stress edge" in the case of a Schwarzschild metric. 
The resulting lightcurve exhibits a modulation similar to that 
observed during a recent X-ray flare from Sgr A*.
\end{abstract}

\keywords{accretion---black hole physics---Galaxy:
  center---magnetohydro\-dynamics---plasmas---Instabilities} 

\section{Introduction}
\label{intro}
Sgr A*'s time-averaged spectrum is roughly a power law below 100\ GHz, 
with a flux density $S_\nu\propto\nu^\alpha$, where $\alpha\sim$\ 0.19--0.34. 
In the mm/sub-mm region, however, Sgr A*'s spectrum is dominated by a
``bump'' \citep{Zylka92}, indicative of two different emission components 
\citep{Melia00,Agol00}. Higher frequencies correspond to smaller spatial 
scales \citep{Melia92,Narayan95}, so the mm/sub-mm radiation is likely 
produced near the black hole (BH).  X-ray flares detected from Sgr A* 
\citep{Baganoff01,Goldwurm03,Porquet03,Belanger05} may also have been 
produced within this compact region, either from a sudden increase in 
accretion accompanied by a reduction in the anomalous viscosity, or from 
the quick acceleration of electrons near the BH \citep{LiuMelia02,LiuPetrosian04}. 
The energized electrons may also manifest themselves via enhanced emission 
in a hypothesized jet \citep{markoff01}.

Near-IR flares detected from Sgr A* appear to be modulated
with a variable period $\approx 17$ minutes 
\citep{Genzel03,Eckart06,Mayer06,Eckart07}. The X-ray
and near-IR flares may be coupled via the same electron population,
so one may expect similarities in their lightcurves. A long 
X-ray flare detected with XMM-{\it Newton} in 2004 also appears to have a 
modulated lightcurve, though not characterized by a fixed period 
\citep{Belanger08}. If real, the modulation in both the near-IR and 
X-ray events is almost certainly quasi-periodic rather than periodic, 
with a decreasing cycle from start to end.
But are the fluctuations due to a single azimuthal perturbation 
(i.e., a ``hotspot"), or from a global pattern of disturbance 
with a speed not directly associated with the underlying Keplerian 
period \citep[][]{TM06,Fal07}? In this 
{\it Letter} we examine the nature of the observerd quasi-period, 
and focus on its implications for the flow of matter through the
innermost stable circular orbit (ISCO). A principal result of this 
study is a ray-tracing simulation of the general-relativistically (GR)
modulated lightcurve produced as the disrupted plasma spirals 
inwards towards the disk's ``stress edge" \citep{Krolik02}.

\section{Background}
\label{back}
Magnetohydrodynamic (MHD) simulations of Sgr A*'s disk 
demonstrate the growth of a Rossby-wave instability, enhancing the
accretion rate for several hours, possibly accounting for the observed
flares \citep{TM06}. The lightcurve produced by GR effects during a 
Rossby-wave induced spiral pattern in the disk fit the data relatively 
well, with a quasi-period associated with the pattern speed rather than
the Keplerian motion \citep{Fal07}. However, MHD simulations of 
black-hole accretion suggest that magnetic reconnection might take 
place within the plunging region, due to the presence of a
non-axisymmetric spiral density structure, initially caused by the
magnetorotational instability (MRI) associated with differential
rotation of frozen-in plasma \citep[see, e.g.,][]{Hawley01}.  

In this case, the accreting flow is no longer Keplerian because of a
radial velocity component. If Sgr A*'s quasi-period of $\sim$ 17--25 minutes is 
associated with this kind of process rather than a pattern rotation, 
it would place the corresponding emission region at $0.73$--$0.94\;
r_{\rm ISCO}$ radii, below the ISCO (where $r_{\rm ISCO}= 3r_{s}=
6GM/c^2$) for a Schwarzschild BH. Theoretically, we may therefore 
distinguish the ISCO from the radius at which the inspiraling material 
actually detaches from the rest of the magnetized disk---the so-called 
{\it stress} edge \citep{Krolik02}. The X-ray modulation would then be 
associated with the ever-shrinking period of the emitting plasma as 
it spirals inwards from the magnetic flare. 

Interest in ``hotspots'' began in the early 1980's in connection with 
quasi-periodic flux modulations observed in BHs accreting from a binary 
companion. The hotspots are possibly overdense 
emission regions associated with magnetic instabilities. But even with a 
hotspot, a Newtonian disk does not produce a modulation since its aspect 
does not affect the total luminosity observed from it. Other than a dynamical 
periodicity (such as that due to an azimuthal, radial, or vertical oscillation), 
only GR effects can produce time-dependent photon trajectories resulting in 
a modulated lightcurve \citep[see e.g.][]{cb73,abramo91,kb92,Holly95,Falcke00,Brom01}. 
Even so, the ``standard" disk picture of hotspot modulation has been based on 
Keplerian motion, for which one then expects a time variability directly related 
to the Keplerian frequency. Here, the modulation is not associated with 
such a fixed Keplerian frequency, but from a shrinking orbit and 
a monotonically decreasing period (see \S\ \ref{model}). The relevance 
of hotspots has already appeared in \citep[][for review]{Holly95,Mayer06,
Eckart07,Melia07}. What is lacking, however, is a non Keplerian treatment 
of the motion with the intent of probing the stress edge itself.

So where exactly is the inner edge of the accretion disk in
Sgr A*? This is a question asked in a broader context by
\citet{Krolik02}, whose MHD simulations 
of the plunging region in a pseudo-Newtonian potential 
identified several characteristic inner radii. Here, we assume a 
non-spinning BH, so our model pertains solely to the Schwarzschild case.

The monotonic decrease of the period during the flares suggests that
we are witnessing the evolution of an event moving inwards across 
the ISCO. The inflow time scale, $t_{\rm inflow}$, which determines
the rate at which plasma can move from one orbit to another, is given
by $\tau_v=r_g/v_{\rm inflow} \approx 9.6\,(r/r_g)^{1/2}$ minutes
\citep{LiuMelia02} and is approximately 23.5 minutes at $r=3r_s=6r_g$,
corresponding to the ISCO for a non-rotating (i.e., $a/r_g=0$)
BH. This time scale does not explicitly depend on a viscosity parameter 
since the viscosity is directly tied to the MRI physical process via
the induced Maxwell stress \citep{LiuMelia02}. The inflow time scale
defined here characterizes local processes occurring within the innermost
portion of the disk during the flares. By comparison, the dynamical
time scale, $t_d\approx 1.3\,(r/r_g)^{3/2}$, is roughly $19$ minutes
at this radius \citep{LiuMelia02}. Thus, the azimuthal asymmetry
giving rise to the modulated flux during the flare may be due to a
transient event associated with either a dynamical or viscous process
close to the ISCO \citep{Melia01a}. 

For a BH mass of $3.6\times 10^6$ M$_\odot$, the inflow time scale
at $r\approx 2.5\,r_s$ (inferred from the {\it average} period) 
is just slightly larger than the average period, so the event 
could be due to the sudden reconfiguration of magnetic field lines 
frozen into plasma rapidly approaching the ISCO and then 
flowing across it towards the event horizon. Matter flowing past the 
ISCO may still remain ``magnetically" coupled to the outer accretion 
flow, so a dynamically more meaningful 
radius is the so-called {\it stress edge}, where plunging matter loses 
dynamical contact with the material farther out \citep{Krolik02}. 
This may simply be defined as the surface on which the inflow speed 
first exceeds the magnetosonic speed. 

\begin{figure}[ht]
\begin{center}
\epsscale{1.0}
\includegraphics[scale=0.3,angle=-90]{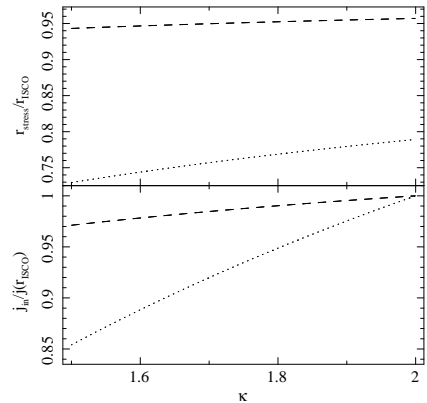}
\caption{\footnotesize Upper panel: The stress edge radius, $r_{\rm
stress}$, in units of $r_{\rm ISCO}$, as a function of $\kappa$, 
the exponent in the power-law formulation of $\Omega(r)$. The dotted and 
dashed curves represent a period of 17 and 25 minutes, respectively,
using a black-hole mass of $3.6\times 10^6$ M$_\odot$ \citep{Schodel03}. 
Lower panel: The corresponding ratio of accreted specific angular momentum, 
$j_{\rm in}$, to the specific angular momentum at the ISCO. 
\label{fig:fig1}}
\end{center}
\end{figure} 

In their simulations, 
\citet{Krolik02} determined that this surface occurs somewhere 
between $0.77r_{\rm ISCO}$ and $r_{\rm ISCO}$. The specific angular 
momentum $j=r^{2}\Omega(r)$, in terms of the orbital angular frequency
$\Omega(r)$, continues to fall below $r_{\rm ISCO}$, though $\Omega$ may
not necessarily trace its Keplerian value, $\Omega_K(r)\equiv
(GM/r^3)^{1/2}$. In the absence of any magnetic
coupling across $r_{\rm ISCO}$, matter would retain all of its specific
angular momentum at the ISCO, so that the accreted value of $j$, which
we will call $j_{\rm in}$, would then simply be $j_{\rm in}=
r_{\rm ISCO}^2\,\Omega_K(r_{\rm ISCO})$. Instead, the MHD simulations
show that $j_{\rm in}\approx 0.95\,j(r_{\rm ISCO})$, for which 
$r_{\rm stress}$ is then $\sim 0.8\,r_{\rm ISCO}$,
within the range of values indicated by the location of the
trans-magnetosonic surface. 

If the period in Sgr A* is decreasing monotonically, $j(r)$ will not
follow its Keplerian value below $r_{\rm ISCO}$. Therefore we 
will adopt the formulation $\Omega(r)=\Omega_0\,r^{-\kappa}$ to fit
the data in \S\ 3. Clearly, $\kappa=3/2$ corresponds to 
Keplerian rotation; $\kappa$ is $2$ in the extreme case of angular 
momentum conservation. A reasonable fit to the data would therefore 
be associated with $3/2\le\kappa\le 2$. At the boundary $r_{\rm ISCO}$, 
we expect $\Omega=\Omega_K$, which 
then forces the constant $\Omega_0$ to have the value $c\sqrt{r_{g}}\,
r_{\rm ISCO}^{\kappa-3/2}$. We calculate $r_{\rm stress}$ using the 
quasi-periods 17 and 25 minutes emerging from the X-ray lightcurve
(see \S\ \ref{model}), and this is plotted as a function of $\kappa$ 
in Fig.~\ref{fig:fig1}.
The radius $r_{\rm stress}$ falls within the range $0.73$--$0.96\; 
r_{\rm ISCO}$ for all permitted values of $\kappa$. The corresponding 
accreted specific angular momentum, for the same parameters as used 
before (see Fig.~\ref{fig:fig1}), is $0.85\,j(r_{\rm ISCO})\le j_{\rm in}
\le j(r_{\rm ISCO})$ as a function of $\kappa$. The ratio 
$j_{\rm in}/j(r_{\rm ISCO})=0.95$ from the MHD simulations 
would require $\kappa\sim 1.8$, for which $r_{\rm stress} 
\sim 0.77\, r_{\rm ISCO}$. These results are consistent with the 
MHD simulations, indicating that the infalling plasma below the ISCO 
remains magnetically coupled to the outer disk, though the dissipation 
of angular momentum is not quite strong enough in this region to force 
the gas into Keplerian rotation.

\section{The Inspiraling Plasma Model}
\label{model}
\begin{figure}[ht]
\begin{center}
\epsscale{1.0}
\includegraphics[scale=0.45,angle=-90]{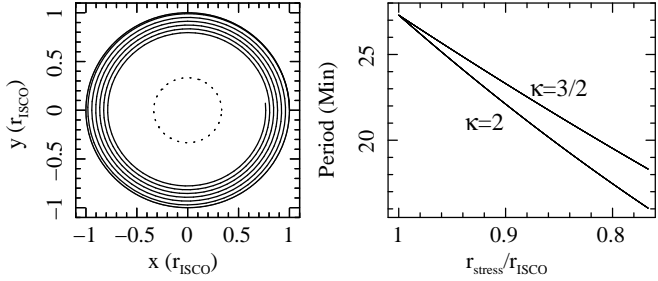}
\caption{\footnotesize Upper panel: The inspiraling trajectory
of the hotspot, beginning at $r_{\rm ISCO}$ and terminating at
$0.74r_{\rm ISCO}$. The dotted circle represents the location of
the event horizon. Lower panel: The period as a function of
the stress radius for the two extreme values of $\kappa$ adopted
here, assuming a black-hole mass of $3.6\times10^6\;M_\odot$.
\label{fig:fig2}}
\end{center}
\end{figure}

With $\Omega(r)$ known, we now incorporate strong gravitational 
effects in a geometrically and optically thin disk, describing 
the inspiraling disturbance using coordinates ($r,\theta,\varphi$)
in the co-rotating frame. In Fig.~\ref{fig:fig2}, we show the inspiraling 
trajectory and duration of the emitting plasma. The 
observer is located at infinity with a viewing angle {\it i} 
relative to the $z'$-axis in the non-rotating frame, at (observer) 
polar coordinates ($r',\theta',\varphi'$). The deflection angle of 
a photon emitted by plasma in the inspiraling region is $\psi$, 
varying periodically with $\cos\, \psi = \cos\,i\, \cos \,\varphi$,
for a disk in the plane $\theta=\pi/2$. Also, for $G=c=1$, 
the BH's horizon occurs at $r_{s} = 2M$, and the last stable 
orbit is located at $r_{\rm ISCO} = 3r_{\rm s}$. 

\vskip -0.1in

We calculate the lightcurve using a full ray-tracing algorithm
\citep[see][]{Luminet79,Fal07}. 
The disk from $r_{\rm ISCO}$ to 90$r_{s}$ is 
an unperturbed, Keplerian flow, with angular velocity $\Omega_{\rm
  K}$, and with specific angular momentum $j_{\rm K}=
r^{2}u^{\varphi}/u^{t}=r^{2}\Omega_{\rm K}$.
The corresponding four-velocity of the effective flow is then
$(u^{t},u^{r},u^{\theta},u^{\varphi})=u^{t}(1,0,0,\Omega_{K})$, 
where $u^{t}=(1-3M/r)^{-1/2}$ \citep{MTW73}. The accretion 
flow is no longer Keplerian below the ISCO. 

Triggering a perturbation induces an azimuthal asymmetry in the region 
$r_{\rm stress} \approx 0.73 <r< 0.9 r_{\rm ISCO}$. Below $r_{\rm ISCO}$, 
we use a simple representation of the bulk velocity field, in which
$\Omega(r)=u^{\varphi}_{\rm sw}/u_{\rm sw}^{t}$, as described e.g., 
in \citet{Fukumura04}:
\vskip -0.1in
\begin{equation}
v^{r}_{\rm sw} = - A_{r} e^{-(r-r_{\rm
    stress})/\Delta_{\rm sw}}\sin^{\gamma_{0}}[k_{r}(r-r_{\rm
    stress})+m\varphi/2-\varphi_{\rm sw}/2]\;.  
\end{equation}
\vskip 0.01in\noindent 
In this case, the specific angular momentum is $j_{\rm in}=r^{2}u^{\varphi}_{\rm
  sw}/u_{\rm sw}^{t}=\Omega_{0} r^{2-\kappa}$.  The subscript ``sw" denotes the
spiral wave, and the number $m$ is the azimuthal wavenumber, fixed to 
be $m=1$ for a single-armed spiral wave. The constant $\gamma_{0}=2$ is 
the width of the spiral wave, $A_{r}=0.1$ and
$A_{\varphi}=0.1$ are the amplitudes chosen to be relatively small, 
$k_{r}$ characterizes a tightness (i.e., the
number of windings) of the spiral, and the effective radial range of
the spiral motion is controlled by $\Delta_{\rm sw}=30$, and
$\varphi_{\rm sw}=0$ denotes the phase of the spiral. Since 
($u^{r}_{\rm sw}$, $u^{\varphi}_{\rm sw}$) is not axisymmetric, the
net velocity field is also non-axisymmetric. For the effective flow then,
$(u_{\rm sw}^{t},u^{r}_{\rm sw},u_{\rm sw}^{\theta},u^{\varphi}_{\rm sw}) =
u_{\rm sw}^{t}(1,v^{r}_{\rm sw},0,\Omega_0\,r^{-k})$, where $u_{\rm sw}^{t}=[
(1-2/m)-(1-2m/r)^{-1}(v_{\rm sw}^{r})^{2}-r^{2}\Omega]^{1/2}$,
corresponding to the four-vector normalization condition
$g_{\alpha,\beta}u^{\alpha}u^{\beta}=-1$.

We consider four GR effects: (i) light-bending, (ii) gravitational 
Doppler effect defined as (1+z), taking into account the non-axisymmetric 
radial and azimuthal components below $r_{\rm ISCO}$, (iii) gravitational
lensing, $d\Omega_{\rm obs}=b\,db\,d\varphi/D^2$ (with $D$ the distance 
to the source), expressed through the impact parameter, and (iv) the travel 
time delay. The relative time delay between photons arriving at the observer 
from different parts of the disk are calculated from the geodesic equation. 
The first photon arrives from phase $\varphi=0$ and $r=r_{\rm ISCO}$, and
defines the reference time, $T_{0}$, which is set to zero. The observed 
time is then the orbital time plus the light-bending travel time delays, 
i.e., $T_{\rm obs}(\varphi_{\rm sw},r,i) = \Omega^{-1}(r)\varphi_{\rm sw} 
+ \Delta T_{\rm GR}$.  

The observed flux at energy $E'$ is $F_{\rm obs}(E') =
I_{\rm obs}(E')d\Omega_{\rm obs}$, where $I_{\rm obs}(E')$ 
is the radiation intensity observed at infinity and
$d\Omega_{\rm obs}$ is the solid angle on the observer's sky 
including relativistic effects. Using the relation 
$I_{\rm obs}(E',\alpha')=(1+z)^{-3} I_{\rm em}(E,\alpha)$, 
a Lorentz invariant quantity that is constant along null geodesics in
vacuum, the intensity of a light source integrated over its effective 
energy range is proportional to the fourth power of the redshift
factor, $I_{\rm obs}(\alpha')= (1+z)^{-4}I_{\rm em}(r,\varphi)$, 
$I_{\rm em}(r,\varphi)$ being the intensity measured in the rest 
frame of the inspiraling disturbance \citep{MTW73}. The 
disk radiates an inverse Compton spectrum, $I_{\rm em}$, 
calculated using the parameter scalings, rather than their absolute 
values. The spectrum parameters are 
\citep{Melia01} the disk temperature, $T(r)$, the electron
number density, $n_e(r)$, the magnetic field, $B(r)$, and the disk
height $H(r)$. This procedure gives correct amplitudes in the
lightcurve, though not the absolute value of the flux per se.

The synchrotron emissivity is therefore $j_s \propto B\,n_{\rm nt} 
\propto B\,T\,n_e$, where the nonthermal particle energy is roughly 
in equipartition with the thermal. The X-rays are produced via inverse 
Compton scattering from the  seed
photon number flux. Thus, with $L_{\rm seed}\propto r^3\, j_s$, where $j_s$
is the synchrotron emissivity in units of energy per unit volume per unit
time, the soft photon flux scales as the emitted power divided by the
characteristic area. That is, $F_{\rm seed}\propto r^3\, j_s / r^2 = r j_s$,
which is going to be roughly the same scaling as the seed photon density, so
$n_{\rm seed} \propto r j_s \propto r\,B\,T\,n_e$.  The inverse
Compton scattering emissivity is therefore $j_{ic}  \propto n_{\rm nt}\,
n_{\rm seed} \propto (T\,n_e)^2\,r\,B$.  Thus, $j_{x}\sim j_{ic}$,
and the surface intensity is $I_{\rm em} \propto \int j_{x} ds \propto
j_{x} H$, which gives finally $I_{\rm em} \propto
(T\,n_e)^2\,r\,B\,H$. 

\begin{figure}[ht]
\begin{center}
\epsscale{1.0}
\includegraphics[scale=0.4,angle=-90]{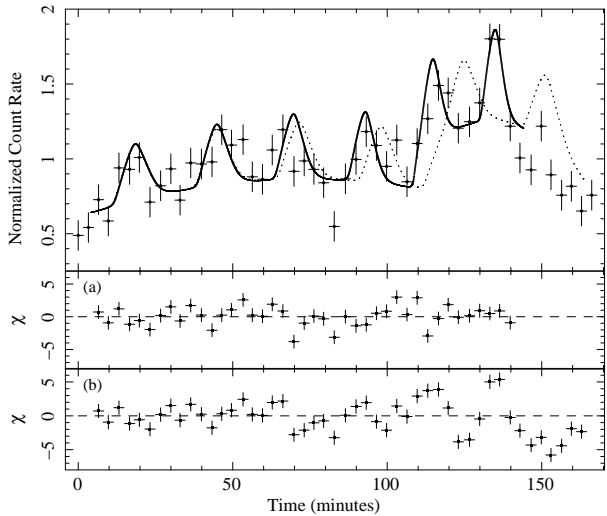}
\caption{\footnotesize Lightcurve of the August 31, 2004
flare in the 2--10 keV energy band \citep{Belanger08}, 
normalized with the observed mean count rate of 0.231 cts s$^-1$ for the
flare duration. The best fit model for an inspiraling disturbance 
is shown by the solid line using an inclination angle 72$^{\circ}$ and 
$\kappa=1.7$. The dotted curve represents a constant Keplerian period 
at the last stable orbit, i.e., $r_{\rm ISCO}$, $i= 72^{\circ}$, and 
$\kappa=1.5$. Panels (a) and (b) show the residuals (in units of sigma)
of the inspiraling and constant-period model, respectively, compared 
to the data.
\label{fig:fig3}}
\end{center}
\end{figure}

The flux at a given azimuthal angle $\varphi$ and radius $r$ is calculated
from a numerical computation of $\psi(\alpha)$, followed by a calculation 
of the Doppler shift, lensing effects, and the flux $F_{\rm obs}$ as a
function of the arrival time. For the persistent emission we use 
the best fit spectral parameters to the
{\it Chandra} data  \citep{Melia01,Baganoff01}, described above as
a surface emissivity $I_{\rm em}$. The observed flare 
normalized flux is modeled with two polynomials, one between 0--100
minutes and the second from 100-160 minutes \citep[see also][]{Mayer06}. 
The value $k_{r}$ is fixed at 11 to have the six observed cycles 
(see Fig. \ref{fig:fig3}, solid line). The free parameters to
fit the data are the inclination angle $i$ and the $\kappa$ 
value. The integrated flux is calculated for an extended spiral 
wave 90$^{\circ}$ long in the azimuthal direction and  $\Delta r =
0.28r_{g}$ in the radial direction, plus the persistent emission. 
The MHD simulations show that in the innermost part of the
disk a spiral-arm often expands out to $\sim90^{\circ}$ (see, e.g.,
Hawley 2001). The radial extent of the inspiraling region is set
by the observed condition that six cycles 
should fit within the overall migration of the plasma from the ISCO 
to the stress edge. In Fig.~\ref{fig:fig3} (solid line), we show the 
best fit model for $72\pm3^{\circ}$ and $\kappa=1.7\pm0.05$.   

\section{Conclusion}
\label{conclusion}

If we adopt the simple view that the last period 
corresponds to the ISCO, then Sgr A* with a mass of $3.6\times 10^6\;
M_\odot$ must be spinning at a rate $a/r_g\gtrsim 0.2-0.4$. With a more 
realistic analysis of the magnetic coupling between matter in the 
plunging region and that beyond the ISCO, we conclude that the peak 
of the instability probably occurs at $\sim 0.97r_{\rm ISCO}$, where 
the period is $\sim 25$ minutes, and the flaring activity continues 
as the plasma spirals inwards, ending several orbits later when the 
matter crosses the stress edge at $\sim 0.8 r_{\rm ISCO}$. 

The significance of the fit for an inspiraling disturbance 
is $\chi^2/d.o.f. = 92.4/39$, compared to $\chi^2/d.o.f. = 285.2/46$
for a fixed Keplerian period (see dotted curve in 
Figure~\ref{fig:fig3}). An inspiraling disturbance 
is preferred over a fixed orbit by a factor 2.6 in the reduced $\chi^2$.
The residuals in the lower panels of Figure ~\ref{fig:fig3} show
that the model using a fixed period produces modulations that are 
progressively shifted in phase with respect to the data, by as much as
$\sim 16.5$ minutes by the end of the flare. The inspiraling
model, on the other hand, follows the evolution of the flare
and therefore fits the data much better. Plasma on such an orbit 
also produces a constant pulsed fraction$=(I_{\rm max}-I_{\rm mean})/
(I_{\rm max}+I_{\rm min})$ of $\sim 9\%$, compared with a linear
increase from $\sim9\%$ to $\sim 11\%$ for the inspiralling wave; 
this effect is due to a radially-dependent gravitational 
lensing effect. Together, these two effects render the inspiraling 
scenario a better explanation for the data than the fixed orbit 
disturbance. 

\acknowledgments
\noindent MF is grateful to Keigo Fukumura for helpful discussions. This 
research was supported by NSF grant AST-0402502 in Arizona,
and by the French Space Agency (CNES).

\end{document}